\documentclass[12pt,a4paper]{article}
\pdfoutput=1
\usepackage{amsmath}
\usepackage{amssymb,exscale}
\usepackage{amsfonts,amsthm}
\usepackage{graphicx, float, epsfig}
\usepackage{mathrsfs}
\usepackage{slashed}
\usepackage{textcomp}
\usepackage[T1]{fontenc}
\usepackage[utf8]{inputenc}
\usepackage[table]{xcolor}
\usepackage[lowtilde]{url}
\usepackage{cancel}
\numberwithin{equation}{section}
\usepackage{verbatim} 
\usepackage{subcaption}
\usepackage{appendix}
\usepackage{hyperref}

\newcommand{\be}{\begin{equation}}
\newcommand{\ee}{\end{equation}}
\newcommand{\ba}{\begin{eqnarray}}
\newcommand{\ea}{\end{eqnarray}}
\newcommand{\bay}{\begin{array}{rcl}}
\newcommand{\eay}{\end{array}}
\newcommand{\ra}{\rightarrow}

\newcommand{\ri}{{\rm i}}

\begin{document}

\title{Particles, Forces and the Early Universe}

{\footnotesize
\author{Risto Raitio\footnote{~E-mail: risto.raitio@gmail.com}\\	
Helsinki Institute of Physics, P.O. Box 64, \\00014 University of Helsinki, Finland}

\date{July 10, 2025} 
\maketitle 

\abstract{\noindent
This article reveals the unexpected result that three a priori distinct ideas—global supersymmetry of preons, Hartle-Hawking cosmology, and Chern-Simons quantum gravity—share common concepts that offer paths beyond the Standard Model. Differences from the MSSM are discussed.}

\vskip 1.0cm
\noindent


\vskip2.0cm
\noindent
Published in Nuclear Physics B, 116990 (2025). \\
\href{https://doi.org/10.1016/j.nuclphysb.2025.116990}{10.1016/j.nuclphysb.2025.116990}

\newpage

\tableofcontents
\vskip 1.0cm

\section{Introduction}
\label{intro}

At first glance, there appears to be no connection between the three distinct ideas considered in this article: 
(1) supersymmetric preons as fundamental particles,
(2) Hartle-Hawking no-boundary cosmological condition, and 
(3) all order finite 3D quantum gravity.

We briefly outline these points. We propose that below the quark-lepton level length scale, $\sim 10^{-18}$ m, there is a topological level of supersymmetric (SUSY) preons. The binding force between preons is a 3D Chern-Simons interaction, which is engineered stronger than Coulomb repulsion between like charge preons. 

To remove the classical initial singularity of the quantum universe, the initial state of the universe must be non-singular. We choose the Hartle-Hawking no-boundary condition for the wave function of the universe. 

Gravitation is based on a recent non-perturbative and all-order perturbatively calculable Chern-Simons (CS) quantum gravity model. The parity violating CS interaction leads to different intensities to polarization states of gravitational waves (GW) provided the source is symmetric. These waves may come from times as early as before inflation started, $t \lessapprox 10^{-35}$s. 

This article discloses a scenario that encompasses the three points listed above.
We propose solutions for some of shortcomings of the Standard Model. Furthermore, we give arguments to support the Fayet conjecture "$Matter \leftarrow\rightarrow Forces$" \cite{Fayet:2016}. 
 
This article is organized as follows. 
Our preon model for visible matter and the dark sector is summarized in section \ref{part}.\footnote{~Section \ref{part} is based on this authors work. Sections \ref{prebin}, \ref{nobowf} and \ref{qg} are from the authors cited.} 
In section \ref{nobowf} we briefly review the Hartle-Hawking no-boundary wave function and Wheeler-DeWitt equation. 
Three-dimensional Chern-Simons gravity with calculational capability is introduced in section \ref{qg} for the very early quantum universe. 
Finally, concluding remarks are presented in section \ref{conout}. 

This note is intended as a brief introduction to selected concepts in particle models and quantum gravity phenomenology. Readers interested in the technical details are encouraged to consult the references.  Many details remain to be determined.

\section{Particle Model}
\label{part}

\subsection{Preons and Particles}
\label{prepar}

The fundamental particles—preons—are organized into vector and chiral supermultiplets \cite{Rai:2018}. Preons are free particles above the energy scale $\Lambda_{cr}$, numerically about $\sim 10^{10} - 10^{16}$ GeV. It is close to the reheating scale $T_R$ and the grand unified theory (GUT) scale. A binding mechanism for the preon bound states has been constructed using the spontaneously broken 3D Chern-Simons theory \cite{Beli:2003} in subsection \ref{prebin}. At $\Lambda_{cr}$ preons form composite states by an attractive Chern-Simons model interaction into of Standard Model quarks and leptons, with the usual SM gauge interactions. Preons have undergone "second quarkization". 

To make the preon scenario compatible with the SM we considered originally, following \cite{wz:1974}, the following Lagrangians \ref{chiralcharge} and \ref{chiralcolor}. To include charged matter we define the charged chiral field Lagrangian for fermion $m^-$, complex scalar $s^-$ and the electromagnetic field tensor $F_{\mu\nu}$\footnote{~The next two equations are in standard 4D form. They are not used quantitatively below.}
\be
\mathcal{L}_{QED} = -\frac{1}{2}\bar{m}^- \gamma^{\mu}(\partial_{\mu} +ieA_{\mu})m^- - \frac{1}{2} (\partial s^-)^2 - \frac{1}{4}F_{\mu\nu}F^{\mu\nu}~.
\label{chiralcharge}
\ee

We assign color to the neutral fermion $m \ra m^0_i$ ($i = R, G, B$). The color sector Lagrangian is then
\be
\mathcal{L}_{QCD} = -\frac{1}{2}                                                                        \sum_{{i=R,G,B}}\Big[\bar{m}^0_i \gamma^{\mu}(\partial_{\mu} +igG^a_{\mu}t_a)m^0_i  \Big] - \frac{1}{4}G_{\mu\nu}^aG^{\mu\nu}_a ~.
\label{chiralcolor}
\ee

With the above Lagrangians $\mathcal{L}_{QED}$ and $\mathcal{L}_{QCD}$ in mind we now define a sample model for preons by the supermultiplets shown in table \ref{tab:table1}.\footnote{~The indices of particles in tables \ref{tab:table1} and \ref{tab:table2} are corrected from those in \cite{Rai:2018, Rai:2023}.}

The superpartners of standard model particles are formed of $s^-$ and $\sigma^0_i$ composites. They generate a rich spectroscopy with lowest composite state masses in the usual lepton/hadron mass scale. Therefore they should be detectable with present accelerator experiments. This challenge remains unresolved, as in the MSSM. The dark sector is obtained from the scalar $\sigma^0_R\sigma^0_G\sigma^0_B$ and the axion multiplet \{$a, n$\} in table \ref{tab:table1} (if the axion(s) are found).

In the MSSM, baryon number and lepton number are no longer conserved by all of the renormalizable couplings. But baryon number and lepton number conservation have been tested very precisely. These couplings have to be very small not to be in conflict with experimental data. Denote baryon number by $B$, lepton number by $L$ and spin by $s$, then R-parity $P_R = (-1)^{(3B-L)+2s}$ is a symmetry that forbids these couplings. All SM particles have R-parity of +1 while superpartners have R-parity of -1.

In the preon model, $B = L = 0$. This leads to a situation where a group of preons and antipreons can form either hydrogen or antihydrogen atoms in the after preons have formed quarks and leptons. Statistical fluctuations cause $N_H \neq N_{\bar{H}}$. This creates the numerically small baryon asymmetry $n_B/n_{\gamma}$ as discussed in \ref{barasym}.

\begin{table}
	\begin{center}
		\captionsetup{width=.8\linewidth}
		\begin{tabular}{|l|l|} 
			\hline
			Multiplet & Particle, Sparticle \\ 
			\hline   
			chiral multiplets spins 1/2, 0 & $m^-, ~s^-; ~m^0_i,~\sigma^0_i$~; $n, a$ \\
			vector multiplets spins 1, 1/2 & $\gamma, m^0; ~g_i , m^0_i$ \\  
			\hline
		\end{tabular}
		\caption{\footnotesize The particles $m^-, m^0$ are charged and neutral, respectively, Dirac spinors. The particle $s^-$ is a charged scalar particle. The $a$ is axion and $n$ axino \cite{Pe:1977, Wei:1978, Wil:1978}. $m^0$ is color singlet particle and $\gamma$ is the photon. $m_i$ and $g_i, \sigma^0_i$ (i = R, G, B) are zero charge color triplet fermions and bosons, respectively.}
		\label{tab:table1}
	\end{center}
\end{table}   

The matter-preon correspondence for the first two flavors (r = 1, 2; i.e., the first generation) is indicated in table \ref{tab:table2} for the left-handed particles. 

\begin{table}
	\begin{center}
		\captionsetup{width=.8\linewidth}
		\begin{tabular}{|l|l|} 
			\hline
			SM Matter 1st gen. & Preon state \\ 
			\hline                                       
			$\nu_e$ & $m^0_R m^0_G m^0_B$ \\     
			$u_R$ & $m^+ m^+ m^0_R$ \\			  
			$u_G$ & $m^+ m^+ m^0_G$ \\  			
			$u_B$ & $m^+ m^+ m^0_B$ \\							         
			$d_R$ & $m^- m^0_G m^0_B$ \\		 
			$d_G$ & $m^- m^0_B m^0_R$ \\			
			$d_B$ & $m^- m^0_R m^0_G$ \\
			$e^-$ & $m^- m^-m^-$ \\	
			\hline
			Sfermions & Preon state   \\
			\hline
			$\tilde{\nu}$ & $\sigma^0_R\sigma^0_G\sigma^0_B$ \\
			$\tilde{u}_R$ & $s^+s^+\sigma^0_R$  \\
			$\tilde{u}_G$ & $s^+s^+\sigma^0_G$  \\
			$\tilde{u}_B$ & $s^+s^+\sigma^0_B$  \\
			$\tilde{d}_R$ & $s^-\sigma^0_G\sigma^0_B$  \\
			$\tilde{d}_G$ & $s^-\sigma^0_B\sigma^0_R$  \\
			$\tilde{d}_B$ & $s^-\sigma^0_R\sigma^0_G$  \\
			$\tilde{e}^-$ & $s^-s^-s^-$  \\
			\hline
			W-Z Dark Matter & Particle \\
			\hline  
			$\sigma^0_R\sigma^0_G\sigma^0_B$ & dark scalar \\
			boson (or BC) & $s$, axion(s) \\
			$e'$ & axino $n$ \\
			meson, baryon $o$ & $n\bar{n}, 3n$ \\
			nuclei (atoms with $\gamma ')$ & multi $n$ \\
			celestial bodies & any dark stuff \\	 
			black holes & anything (neutral) \\
			\hline
		\end{tabular}
		\caption{\footnotesize Low energy visible and Dark Matter with corresponding particles and preon composites. $m^0_i ~(i = R, G, B)$ is color triplet, $m^{\pm}$ are color singlets of charge $\pm 1/3$. $s^-$ and $\sigma^0_i$ (i = R, G, B) are scalars. Sfermions are indicated by $\tilde{S}$. $e'$ and $\gamma'$ refer to dark electron and dark photon, respectively. BC stands for Bose condensate.}
		\label{tab:table2}
	\end{center}
\end{table}

After quarks are formed by the process described in \cite{Rai:2018} the SM octet of gluons emerges. To make observable color neutral, integer charge states (baryons and mesons) we proceed as follows. The local $SU(3)_{color}$ octet structure is formed by quark-antiquark composite pairs as follows (with only the color charge indicated):
\be
\rm{Gluons}: \footnotesize{R\bar{G}, R\bar{B}, G\bar{R}, G\bar{B}, B\bar{R}, B\bar{G}, \frac{1}{\sqrt{2}}(R\bar{R}-G\bar{G}), \frac{1}{\sqrt{6}}(R\bar{R}+G\bar{G}-2B\bar{B})}~.
\label{gluons} 
\ee

Finally, we briefly and heuristically introduce the weak interaction - the scalar sector is rather complex. For simplicity, we append the Standard Model electroweak interaction in our model as an $SU(2)_Y$ Higgs extension with the weak bosons presented as composite pairs, such as gluons in (\ref{gluons}). 

The Standard Model and dark matter are formed by preon composites in the very early universe at temperature of approximately the reheating value $T_R$. Because of spontaneous symmetry breaking in three-dimensional QED$_3$ by a heavy Higgs-like particle the Chern-Simons action can provide by M\"{o}ller scattering mediated by two particles (the Higgs scalar and the massive gauge field) a binding force stronger than Coulomb repulsion between equal charge preons. The details of preon binding and a mechanism for baryon asymmetry in the universe are presented in subsection \ref{prebin}.  

Chern-Simons theory with larger groups such as $G = U(N_c)$ with fundamental matter and flavor symmetry group $SU(N_f) \times SU(N_f)$ have been studied, for example \cite{Ka:2013}, but they are beyond the scope of this article.

\subsection{Preon Binding}
\label{prebin}

An immediate question for table \ref{tab:table2} particles is the Coulomb repulsion between like charge preons. This problem has been solved for polarized electrons in \cite{Beli:2003}\footnote{We take their low energy result as a first approximation.}
where the authors derived an interaction potential electrons in the framework of a Maxwell-Chern-Simons QED$_{3}$ with spontaneous breaking of local U(1) symmetry. 
An attractive electron-electron interaction potential was found whenever the Higgs sector contribution is stronger than the repulsive contribution of the gauge sector, provided appropriate fitting of the free parameters is made. 

We generalize the results for $e^{-}e^{-}$ binding energy in \cite{Beli:2001, Ferr:2001} for preons. One starts from a QED$_{3}$ Lagrangian built up by two Dirac spinor polarizations, $\psi _{+},\psi _{-})$ with SSB. The authors evaluate the M\"{o}ller scattering amplitudes in the nonrelativistic approximation. The Higgs and the massive photon are the mediators of the corresponding interaction in three different polarization expressions: $V_{_{\uparrow \uparrow }},V_{_{\uparrow \downarrow }},V_{_{\downarrow \downarrow }}$.

The action for a QED$_{3}$ model is built up by the fermionic fields ($\psi _{+},\psi _{-}$), a gauge $\left( A_{\mu }\right) $ and a complex scalar field $\left( \varphi \right)$ with spontaneous breaking of the local U(1)-symmetry \cite{DelC:1998, Beli:2001} is

\begin{align}
	S_{\rm{QED_3-MCS}} & = \int d^{3}x\{-\frac{1}{4}F^{\mu \nu }F_{\mu \nu }+i\overline{
		\psi }_{+}\gamma ^{\mu }D_{\mu }\psi_{+}+i\overline{\psi}_{-}\gamma^{\mu}D_{\mu}\psi 
	_{-}+ \nonumber \\
	& \theta \epsilon ^{\mu v\alpha }A_{\mu }\partial _{v}A_{\alpha }-m_{e}(
	\overline{\psi }_{+}\psi _{+}-\overline{\psi }_{-}\psi_{-})+  \nonumber \\
	& - y(\overline{\psi}_{+}\psi_{+}-\overline{\psi }_{-}\psi_{-})\varphi^{\ast}\varphi +D^{\mu}\varphi^{\ast}D_{\mu}\varphi -V(\varphi^{\ast}\varphi),  
\label{actionMCS}
\end{align}
where $V(\varphi^{\ast}\varphi)$ is the sixth-power $\varphi$ self-interaction potential

\begin{equation}
	V(\varphi ^{\ast }\varphi )=\mu ^{2}\varphi ^{\ast }\varphi +\frac{\zeta }{2}
	(\varphi ^{\ast }\varphi )^{2}+\frac{\lambda }{3}(\varphi ^{\ast }\varphi
	)^{3},
\end{equation}
which is the most general one renormalizable in $1+2$ dimensions \cite{Delcima}. 

In $\left( 1+2\right)$ dimensions, a fermionic field has its spin polarization fixed up by the mass sign \cite{Binegar}. In the action (\ref{actionMCS}) there are two spinor fields of opposite polarization. In this sense, there are two positive-energy spinors, or families, each one with one polarization state according to the sign of the mass parameter.

Considering $\langle \varphi \rangle =v,$ the vacuum expectation value for
the scalar field squared is given by
\[
\langle \varphi ^{\ast }\varphi \rangle =v^{2}=-\zeta /\left( 2\lambda
\right) +\left[ \left( \zeta /\left( 2\lambda \right) \right) ^{2}-\mu
^{2}/\lambda \right] ^{1/2}, 
\]
The condition for minimum is $\mu ^{2}+\frac{\zeta }{2} v^{2}+\lambda v^{4} = 0$. After the spontaneous symmetry breaking, the scalar complex field can be parametrized by $\varphi =v+H+i\theta $, where $H$ represents the Higgs scalar field and $\theta $ the would-be Goldstone boson. To preserve renormalizability of the model, one adds the gauge fixing term $\left( S_{R_{\xi}}^{gt}=\int d^{3}x[-\frac{1}{2\xi }(\partial ^{\mu }A_{\mu }-\sqrt{2}\xi M_{A}\theta )^{2}]\right) $ to the broken action. By keeping only
the bilinear and the Yukawa interaction terms, one has finally

\begin{align}
	{S}_{{\rm {CS-QED_3}}}^{{\rm SSB}} & =\int d^{3}x\biggl\{-\frac{1}{4}F^{\mu \nu}
	F_{\mu \nu }+\frac{1}{2}M_{A}^{2}A^{\mu }A_{\mu } \nonumber \\
	& -\frac{1}{2\xi }(\partial^{\mu }A_{\mu })^{2}+\overline{\psi }_+(i\cancel\partial -m_{eff})\psi _{+} \nonumber \\
	& +\overline{\psi }_{-}(i\cancel\partial +m_{eff})\psi _{-}+ \frac{1}{2}
	\theta \epsilon ^{\mu v\alpha }A_{\mu }\partial _{v}A_{\alpha }  \nonumber \\
	& +\partial ^{\mu }H\partial _{\mu }H-M_{H}^{2}H^{2} +\partial ^{\mu }\theta
	\partial _{\mu }\theta -M_{\theta }^{2}\theta ^{2} \nonumber \\
	& -2yv(\overline{\psi }_{+}\psi _{+}-\overline{\psi }_{-}\psi _{-})H-e_{3}\left( \overline{\psi }
	_{+}\cancel A\psi _{+}+\overline{\psi }_{-}\cancel A\psi _{-}\right) \biggr\}  \label{actionMCS3}
\end{align}
where the mass parameters, 
\begin{equation}
	M_{A}^{2}=2v^{2}e_{3}^{2},\text{ \ \ }m_{eff}=m_{e}+yv^{2},\ \ M_{%
		{\small H}}^{2}=2v^{2}(\zeta +2\lambda v^{2}),\text{ \ }M_{\theta }^{2}=\xi
	M_{A}^{2},
\end{equation}
depend on the SSB mechanism. The Proca mass, $M_{A}^{2}$, represents the mass acquired by the photon through the Higgs mechanism. The Higgs mass, $M_{H}^{2}$, is associated with the real scalar field. The Higgs mechanism causes an effective mass, $m_{eff}$, to the electron. The would-be Goldstone mode, with mass $(M_{\theta}^{2}) $, does not represent a physical excitation. One sees the presence of two photon mass-terms in (\ref{actionMCS3}): the Proca and the topological one. The physical mass of the gauge field will emerge as a function of two mass parameters. 

Electron-electron scattering, the potential must exhibit the combination $(l-\alpha ^{2})^{2}$ for the sake of gauge invariance. In order to ensure the gauge invariance one takes into account the two-photons diagrams, which amounts to adding up to the tree-level potential the quartic order term $\left\{\frac{e^{2}}{2\pi\theta }[1-\theta rK_{1}(\theta r)]\right\} ^{2}$. Now one has the following gauge invariant effective potential \cite{Dobroliubov, Kogan}
\begin{equation}
	V_{{\rm MCS}}(r)=\frac{e^{2}}{2\pi }\left[ 1-\frac{\theta }{m_{e}}\right]
	K_{0}(\theta r)+\frac{1}{m_{e}r^{2}}\left\{ l-\frac{e^{2}}{2\pi \theta }%
	[1-\theta rK_{1}(\theta r)]\right\} ^{2}~.  \label{Vmcs}
\end{equation}
In the expression above, the first term corresponds to the electromagnetic
potential, whereas the last one incorporates the centrifugal barrier $\left(l/mr^{2}\right)$, the Aharonov-Bohm term and the two-photon exchange term. One observes that this procedure becomes necessary when the model is analyzed or defined out of the pertubative limit. 

In search for applications to Condensed Matter Physics, one must require $\theta \ll m_{e}$. The scattering potential (\ref{Vmcs}) is then positive. In our preon scenario we have rather $\theta \gg m_{e}$ and the potential is negative leading to an attractive force of Yukawa type.

The action in (\ref{actionMCS}) is three-dimensional. In a rapidly expanding universe four-dimensional general relativity begins to contribute at or before reheating. Therefore the the Einstein-Hilbert action must be added to (\ref{actionMCS}). The embedding of the CS action into the four-dimensional action is described in subsection \ref{csmgr}.

\section{No-Boundary Wave Function}	
\label{nobowf}

We recap first the ground state and, secondly, the dynamical equation of the wave function of the universe.

\subsection{Ground State of the Universe} 
\label{psi0}

A pedagogic review of the Hartle-Hawking no-boundary concept \cite{Hartle:1983ai} is Lehners' article \cite{Le:2023}, which we follow closely in this section (see also the article \cite{Alex:2020}). In quantum theory, ground states can be defined by solving a proper quantum differential equation or considering a Euclidean path integral. The latter is integrated from configurations of vanishing action in the infinite (Euclidean) past, 
\begin{align}
	\psi_0(x,0) = N\int {\cal D}x \, e^{-\frac{1}{\hbar}I_E[x(\tau)]}\,, 
	\label{eq:ground}
\end{align}
where $N$ is a normalization factor and where Euclidean time $\tau$ is related to physical time via $t=-i\tau$. The Euclidean and Lorentzian actions are related via $I_E=-iS.$ An integral from the infinite Euclidean past defines the ground (vacuum) state of the system, which is taken as the initial state. Furthermore, the replacement $t=-i\tau$ shifts one from quantum oscillatory behavior towards semiclassical physics.

When gravity is switched on, according to Hartle and Hawking \cite{Hartle:1983ai} there are two natural choices (i) Euclidean flat space for scattering amplitudes, and (ii) compact Euclidean metrics. In cosmology one only measures the universe at late (finite) times. More importantly, one does so from the inside of the universe. Clearly, option (ii) is more appropriate for cosmology. One advantage of option (ii) is that there is no need to insert an initial state explicitly. The Euclidean integral takes care of the universe in its ground state.

As discussed in Lehners' review \cite{Le:2023}, the no-boundary proposal assumes a fully quantum view of spacetime: the actual spacetime exists only in interaction with either itself or matter. Our perception of classical spacetime comes from interactions between different constituents and bodies, including ourselves, in the universe.

The arguments of wave function are now three-dimensional spatial slices. The path integral is an amplitude from the initial slice with zero three-dimensional volume, to a final slice with metric $h_{ij},$
\begin{align}
	\Psi_{HH}[h_{ij}] = N \int_{\cal C} {\cal D}g_{\mu\nu} \, e^{-I_E[g_{\mu\nu}]}\,, \label{eq:HHpi}
\end{align}
where the integral is calculated over all (inequivalent) compact metrics ${\cal C}$ and $N$ is a normalization factor. The meaning of this amplitude is for the universe to tunnel from nothing to the final state. The initial state "nothing" contains no space, time or matter. 

The wave function \eqref{eq:HHpi} is real valued but it can lead to the definition of probabilities. The present Lorentzian universe will come out because the saddle points of the path integral \eqref{eq:HHpi} are complex. The big bang singularity is avoided because the initial geometry is Euclidean and the universe shrinks to zero size, i.e. the point universe can be called the "South Pole" (see figure \ref{fig:figure1}). The energy density there has its maximal value. 

\begin{figure}[b]
	\centering
	\captionsetup{width=.8\linewidth}
	\includegraphics[width=8cm]{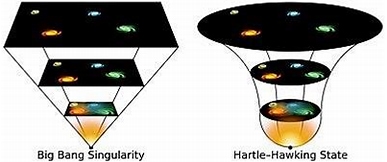}
	\caption{\small Big Bang Universe and Hartle-Hawking Universe.}
	\label{fig:figure1}
\end{figure}

\subsection{Wheeler-DeWitt Equation}	
\label{wdw}

The discussion of subsection \ref{psi0} can also be done using the Hamiltonian form of the action of General Relativity (GR), given by \cite{Arnowitt:1962hi}
\begin{align}
	S=\int \mathrm{d}^3x \ \mathrm{d}t\left[ \dot h_{ij} \pi^{ij}
	-N{\cal H}-N^i{\cal H}_i\right] 
\end{align}
where $\pi^{ij}=\frac{\delta {\cal L}}{\delta \dot{h}^{ij}}=-\frac{\sqrt{h}}{2} \left( K^{ij} - h^{ij}K\right)$  are the momenta conjugate to $h_{ij}$. The Hamiltonian is a sum of
constraints, with lapse $N$ and shift $N^i$ being Lagrange multipliers. The momentum constraint is
\begin{align}
	{\cal H}^i=-2D_j \pi^{ij} +{\cal H}_{matter}^i=0 \,,
\end{align}
and the Hamiltonian constraint
\begin{align}
	{\cal H}=  2 G_{ijkl}\pi^{ij} \pi^{kl}- \frac{1}{2}   \sqrt{h} ({^3}R-2 \Lambda) +{\cal H}_{matter}=0\,,
\end{align}
where $ G_{ijkl} $ is the DeWitt metric \cite{Dewitt:1967yk}
\begin{align}
	G_{ijkl}=\frac{1}{2\sqrt{h}}  \left(h_{ik} h_{jl}+h_{il}
	h_{jk}-h_{ij} h_{kl}\right) \,.
\end{align}

Canonical quantization makes the constraints as operator equations with the familiar substitution
\begin{align}
	\pi^{ij}\to -i\frac{\delta}{\delta h_{ij}}
\end{align}
and correspondingly for matter momenta. We get four equations: the
momentum constraint
\begin{align}
	{\cal H}^i\Psi= 2i D_j \frac{\delta \Psi}{\delta h_{ij} } + {\cal
		H}_{matter}^i \Psi = 0 \,,
\end{align}
and Wheeler-DeWitt equation \cite{Dewitt:1967yk, Dewitt:1967lxx} for the wave function of the universe
\begin{align}
	{\cal H}\Psi (h_{ij},\Phi_{matter}) = \left[ - G_{ijkl} \frac{\delta}{\delta h_{ij} }
	\frac{\delta}{\delta h_{kl} } - \sqrt{h} ({^3}R-2 \Lambda) +{\cal
		H}_{matter} \right] \Psi = 0 \,. \label{eq:wdw}
\end{align}
or
\begin{align}
	\hat{H} \Psi = 0 \rightarrow  \,\, & \hbar^2\frac{\partial^2 \Psi}{\partial q^2} + 12 \pi^4 (\Lambda q - 3)\Psi =0\,.
\end{align}

In the early universe, time is treated as a complex number and it behaves like a spatial dimension. This allows a smooth and finite geometry in all directions of the universe. Imaginary time wipes away the singular boundary of the big bang. 

When the South Pole tunnels into expanding phase, from a plaque around the Pole \footnote{This is where the topological phase may occur.} infinitesimally thin slices of three-dimensional space are formed one after the other transforming the quantum gravity model three-dimensional (to be described in the next section \ref{qg}). The slices expand, trace a 3D space and reach the final, "current" slice occuring in (\ref{eq:HHpi}). At the same time, preon-antipreon pairs are created from the gravitational energy in the neighborhood of the Pole. Preons form (s)fermions due to the spontaneously broken symmetry created, attractive Yukawa-like force described in subsection \ref{prebin}. After inflation comes reheating as in the concordance model. From reheating on, our model will adapts to the standard models of cosmology and particles. Classicalization is obtained in a Wentzel-Kramers-Brillouin (WKB) process (see \cite{Le:2023} for details). Gravity has become classical GR. 

\subsection{Mathematical Supplement}
\label{masu}

We have not yet discussed the possible compatibility of Chern-Simons theory and the Hartle-Hawking no-boundary proposal. Quite promisingly, Kodama showed \cite{Kod:1990} that the Ashtekar-Hamilton-Jacobi equation of General Relativity has the Chern-Simons action as a solution with nonzero cosmological constant. It was therefore expected that when the theory is canonically quantized the quantum constraint equations would have a solution of the form exp$(iS_{CS})$.

The Kodama state has been, however, a subject of debate. In 2003, Witten published a paper arguing that the Kodama state is unphysical, e.g. it has negative energies \cite{Wit:2003}. A few years later, Randono generalized the Kodama state \cite{Rana:2006,Ranb:2006}. He concluded that the Immirzi parameteris generalized to a real value, the theory matches "{\it with black hole entropy, describes parity violation in quantum gravity, and is CPT invariant, and is normalizable, and chiral, consistent with known observations of both gravity and quantum field theory}" \cite{Rana:2006,Ranb:2006}. The physical inner product resembles the MacDowell–Mansouri formulation of gravity, which may include torsion \cite{MacM:1977,Wis:2006}.

Some years after the Kodama paper, Louko \cite{Lou:1995} studied the CS and Hartle-Hawking compatibility problem in more general spacetimes, namely in Bianchi type IX (homogeneous but anisotropic) quantum cosmology with $S^3$ spatial surfaces. He showed that "{\it among the classical solutions generated by $S_{CS}$, there is a two-parameter family of Euclidean space times that have a regular closing of the NUT-type \cite{NUT:1963}. This implies that, in this model, a wave function of the semiclassical form exp$(iS_{CS})$ can be regarded as compatible with the no-boundary proposal of Hartle and Hawking}".

In 2022, Alexander et al. introduced Ashtekar formalism in their approach to quantum gravity \cite{Alex:2022, Asht:1986}. In this formalism the dynamical variables are Yang-Mills gauge field having the SU(2) gauge group. Now the Wheeler-DeWitt equation can be solved exactly. The ground state is the Chern-Simons-Kodama (CSK) state. They "{\it seek to find a new CSK state that includes fermionic matter on the same footing as gravity (...). In this work, we explore a quantization of gravity with the inclusion of fermionic matter by solving both the gravitational and fermionic Hamiltonian constraint. We find an exact wave function that has interesting connections to the CSK state with the inclusion of torsion. We then seek to make contact with the Hartle-Hawking/Vilenkin wave functions of quantum cosmology from this exact wave function.}" \cite{Alex:2022b}.

In 2003, Oda showed \cite{Oda:2003} that "{\it the Kodama state has its origin in topological quantum field theory so that this state has a large gauge symmetry which includes both the usual gauge symmetry and diffeomorphisms. Accordingly, the Kodama state automatically satisfies the quantum Ashtekar constraints."} A related article is \cite{CarT:2005}.

Finally, we mention that Magueijo has shown \cite{Magu:2020} that Chern-Simons wave function is the Fourier dual of the Hartle-Hawking and Vilenkin wave functions.

For now, we set aside the Kodama state, CS-HH compatibility, and other interesting problems, treating sections \ref{nobowf} and \ref{qg} as phenomenological tools for the present. In the next section \ref{csmgr} we recap the classical case.\footnote{~It is possible that quantum gravitational waves are not found.}

\subsection{Baryon asymmetry}
\label{barasym}

We now examine the potential (\ref{Vmcs}) in the early universe. Consider large number of groups of twelve preons each group consisting of four $m^+$, four $m^-$ and four $m^0$ particles. Any bunch of twelve preons (generally 12n, n=1,2, ...) may form only electron and proton (hydrogen atoms H), or only positron and antiproton ($\bar{\rm{H}}$), or some combination of both H and $\bar{\rm{H}}$ atoms (for larger n) \cite{Rai:2023}. This is achieved by arranging the preons appropriately (mod 3) using table \ref{tab:table1}. This way the transition from matter-antimatter symmetric universe to matter-antimatter asymmetric one happens straightforwardly as a statistical effect.

Because the Yukawa force (\ref{Vmcs}) is the strongest force the light $e^-$, $e^+$ and the neutrinos are expected to form first at the very onset of inflation. To obey condition $B-L=0$ of baryon-lepton balance and to sustain charge conservation, for one electron made of three preons, nine other preons have to be created simultaneously, these form a proton. Accordingly for positrons. One neutrino requires a neutron to be created. The $m^0$ carries in addition color enhancing neutrino formation. This makes neutrinos different from other leptons and the quarks. 

Later, when the protons were formed, because preons had the freedom to choose whether they are constituents of $\rm{H}$ or $\bar{\rm{H}}$ there are regions of space of various sizes dominated by $\rm{H}$ or $\bar{\rm{H}}$ atoms. Since the universe is the largest statistical system it is expected that there is only a very slight excesses of $\rm{H}$ atoms (or $\bar{\rm{H}}$ atoms which only means a charge sign redefinition) which remain after the equal amounts of $\rm{H}$ and $\bar{\rm{H}}$ atoms have annihilated. The ratio $n_B/n_{\gamma}$ is thus predicted to be $\ll 1$. 

Fermionic dark matter has in this scenario no mechanism to become "baryon" asymmetric like visible matter. Therefore we expect that part of fermionic dark matter has annihilated into bosonic dark matter. Secondly, we predict there should exist both dark matter and anti-dark matter clumps attracting visible matter in the universe. Collisions of anti-dark matter and dark matter celestial bodies would give us a new source for wide spectrum gravitational wave production (the lunar mass alone is $\sim 10^{49}$ GeV).

\section{Quantum Gravity}
\label{qg}

We first recap quantum gravity, which is expected to be in a major role in the very early universe. We consider a three-dimensional model, which should align with the considerations of the previous section \ref{nobowf}.

\subsection{Wilson Spool}
\label{wlqg}

The recent CS model of quantum gravity by Castro et al. \cite{Ca:2023, Bo:2024} is briefly summarized below. In Euclidean space, fermions $\psi^{\alpha}$ and $\bar \psi^{\alpha}$ are independent. Their transformation properties go under the same representation of the Lorentz group \cite{Ma:2011}. We take the $\gamma_\mu$ matrices to be the hermitian Pauli matrices, and $\gamma_{\mu \nu}$ is defined
\be
\label{dgamma}
\gamma_{\mu \nu} = \frac{1}{2} [\gamma_\mu, \gamma_\nu] =\ri \epsilon_{\mu \nu \rho} \gamma^\rho.
\ee

The three-dimensional Euclidean ${\cal N}=2$ vector superfield $V$ includes the following fields (note that the charge and color indices in table \ref{tab:table1} can be dropped in case of gravity)
\be
V: ~~ A_\mu, \,\, \sigma,\,\, \lambda, \,\, \bar\lambda, \,\, D,
\label{superv}
\ee
where $A_\mu$ is the gauge field, $\sigma$ and $D$ are auxiliary scalar fields, and $\lambda$, $\bar\lambda$ are two-component complex Dirac spinors. The superfield \ref{superv} is as described in \cite{Ma:2011} "{\it the dimensional reduction of the $\mathcal{N}=1$ vector multiplet in four dimensions, and $\sigma$ is the reduction of the fourth component of $A_\mu$.  All fields are valued in Lie algebra $\mathfrak{g}$ of gauge group $G$. For $G=U(N)$ our convention is that $\mathfrak{g}$ is a Hermitian matrix.}" The relevant gauge covariant derivative is then
\be
\partial_\mu + \ri [A_\mu, .. ] .
\ee
The usual gauge field strength is 
\be
F_{\mu \nu}=\partial_\mu A_\nu -\partial_\nu A_\mu + \ri [A_\mu, A_\nu]. 
\ee

The question of gravity-matter coupling was resolved in \cite{Ca:2023a}. Further, the major result of \cite{Ca:2023} is the expression for the partition function (technically, the one-loop determinant) of a massive scalar field minimally coupled to a background metric $g_{\mu\nu}$ \cite{Ca:2023a}
\be
Z_\text{scalar}[g_{\mu\nu}]=\exp\frac{1}{4}\mathbb W[A_L,A_R]~.
\label{eq:spoolsketch}
\ee
Conveniently, it is a gauge invariant object of the Chern-Simons connections $A_{L/R}$.
The object $\mathbb W[A_L,A_R]$, coined the Wilson spool \cite{Bo:2024}, is a collection of Wilson loop operators $W$ \cite{Ca:2023a}
\be
W_R = \rm{Tr}_R \rm{P}exp\Big(i\int_{\gamma}A_{\mu}dx^{\mu}\Big) .
\label{eq:wloop}
\ee
where $\gamma$ is a closed loop in space-time and R is a representation of the gauge group G, wrapped many times around cycles of the base geometry. Supersymmetric localization in the evaluation of Wilson loop expectation values \cite{Ma:2011} with the Wilson spool inserted into the path integral allows a precise and efficient calculation of the quantum gravitational corrections to $Z_{scalar}$ at any order of perturbation theory of Newton’s constant $G_N$. -- More detailed description of Castro et al. \cite{Ca:2023, Bo:2024} is beyond the scope of this note.

Gravitational waves have two polarization states propagating with the speed of light. Parity violation, due to Levi-Civita tensor $\epsilon_{\mu\nu\rho}$ in (\ref{eq:nonacs}), causes the two polarization states to have different intensities.\footnote{~For a discussion of new parity violating interactions, see \cite{Kami:1999}} As to the power spectrum, highest frequency GWs should have strongest intensity due to very early preon-preon (Abelian) M\"{o}ller scattering above energies $\Lambda_{cr} \sim 10^{10} - 10^{16}$ GeV when preons formed a hot gas. All GWs from the time having energy scale $10^{14}$ GeV are maximally redshifted by a factor is approximately $z = \frac{1}{a} - 1 \sim 10^{27}$, which pushes the GW wavelengths beyond the observable range. The NANOGrav detector \cite{nanog} has the low-frequency end at $f \sim 10^{-9} Hz$, corresponding the wavelength of approximately $\lambda_{max}\sim 3\times 10^{17}$ m, or roughly 30–32 light years. Wavelengths larger than $\lambda_{max}$ cause background noise in the detector. LISA detectors \cite{Lisa} in turn are designed to measure GWs in the range of 0.1 mHz – 1 Hz. 

In addition to direct detection of GWs, the next decade CMB data may reveal them through the polarization pattern in the CMB B-modes \cite{hu:2001,doss:2024}, which are difficult to detect. Detection and parametrization of GWs is a subject of itself, see e.g. \cite{Isi:2023}.


\subsection{General Relativity Modified by Chern-Simons Term}
\label{csmgr}

In the following we reiterate subsection \ref{wlqg} for classical astronomical situations including a 3D CS interaction (to be promoted to four dimensions) to General Relativity. This has been studied by Jackiw and Pi \cite{Jack:2003} and Alexander et al. \cite{Alex:2009, Alex:2022c, Alex:2411}. Their modification of GR is the the three-dimensional Chern-Simons term $CS(\Gamma)$ 
\begin{equation}
	CS(\Gamma) = \frac{1}{4 \pi^2} \int d^3 x \  \varepsilon^{ijk} \ (\frac{1}{2}
	\ ^3\Gamma^p_{i q} \, \partial_j  {^3 \Gamma^q_{k p}} + \frac{1}{3}  {^3\Gamma}^p_{iq} \, {^3\Gamma}^q_{j r} \, ^3\Gamma^r_{kp}).
\label{2.1}
\end{equation}
Latin letters range over three values, indexing coordinates on a three manifold. Greek letters denote analogous quantities in four dimensions. The superscript 3 denotes three-dimensional objects.

The Chern-Simons topological current, a four-dimensional quantity, is  
\begin{equation}
K^\mu = 2 \varepsilon^{\mu}  {^{\alpha \beta \gamma}} \left [\frac{1}{2} ~ \Gamma^\sigma_{\alpha \tau} \, \partial_ \beta
~ \Gamma^\tau_{\gamma \sigma} + \frac{1}{3} ~  \Gamma^\sigma_{\alpha \tau} \,  \Gamma^\tau_{\beta \eta} \Gamma^\eta_{\gamma \sigma} \right],
\label{2.4}
\end{equation}
It satisfies the equation
\begin{equation}
	\partial_\mu K^\mu = \frac{1}{2} {^*R}^\sigma_{~\tau} \ {^{\mu \nu}} \ R^\tau {_{\sigma \mu \nu}} \equiv \frac{1}{2} {^*RR},
	\label{2.5}
\end{equation}
where $R^\tau_{~ {\sigma\mu\nu}}$ is the Riemann tensor
\begin{eqnarray}
	R^\tau_{~ {\sigma \mu \nu}} = &\partial_\nu ~ \Gamma_{\mu \sigma} ^\tau  - \partial_\mu ~ \Gamma_{\nu \sigma} ^\tau \negthickspace \negthickspace &+ \Gamma^\tau_{\nu \eta} \Gamma^\eta_{\mu \sigma}
	- \Gamma^\tau_{\mu \eta} \Gamma^\eta_{\nu \sigma},
	\label{2.6} 
\end{eqnarray}
and its dual is $^\ast R^{\tau \ \mu \nu}_{\, \,\sigma}$ 
\begin{equation}
	^*R^{\tau \ \mu \nu}_{\, \,\sigma} = \frac{1}{2} ~ \varepsilon^{\mu \nu \alpha \beta} R^\tau_{~{\sigma \alpha \beta}}.
	\label{2.7}
\end{equation}
Note that the zero component of $K^{\mu}$, i.e. $K^0$, is not related to the Chern-Simons term (\ref{2.1}).\footnote{~Extending the Einstein theory with a Chern-Simons term can be done in different ways.}

We now choose the following Einstein-Hilbert-CS action \cite{Jack:2003}
\begin{equation}
	{\rm I}= \frac{1}{16 \pi G} \int d^4 x \left( \sqrt{-g} R + \frac{1}{4} \theta {^*RR} \right) = \frac{1}{16 \pi G} \int d^4 x \left(\sqrt{-g} R - \frac{1}{2} v_\mu K^\mu \right),
\label{eq:ehac}
\end{equation}
where $\theta$ is the CS coupling field, and $v_\mu \equiv \partial_\mu \theta$ is the embedding coordinate. 
The variation of the first term in the integrand with respect to $g_{\mu \nu}$ produces the usual Einstein tensor $G^{\mu \nu} \equiv R^{\mu \nu} -\frac{1}{2} ~ g^{\mu \nu} R$.
The variation of the second, topological term gives a traceless symmetric, second-rank tensor, which we call the four-dimensional Cotton tensor $C^{\mu\nu}$
\begin{eqnarray}
	C^{\mu \nu}=\frac{-1}{2 \sqrt{-g}} \left[ v_\sigma\left(\varepsilon^{\sigma \mu \alpha \beta} D_\alpha R^\nu_\beta + \varepsilon^{\sigma \nu \alpha \beta}
	D_\alpha R^\mu_\beta \right) + v_{\sigma \tau}\left(^*R^{\tau \mu \sigma \nu} + ^*R^{\tau \nu \sigma \mu} \right)\right].
	\label{2.10}
\end{eqnarray}

The above deformation of Einstein's equation finally reads
\begin{equation}
	G^{\mu \nu} + C^{\mu \nu} = -8 \pi G T^{\mu \nu}.
	\label{2.11}
\end{equation}

For consistency, let us take the covariant divergence of (\ref{2.11}). The Bianchi identity enforces $D_\mu G^{\mu \nu}=0$. In the right hand side, diffeomorphism invariance of matter degrees of freedom implies that $D_\mu T^{\mu \nu} =0$. But the covariant divergence Cotton tensor is non-zero \cite{Jack:2003}
\begin{equation}
	D_\mu C^{\mu \nu} = \frac{1}{8 \sqrt{-g}} v^\nu {^*RR}.
	\label{2.12}
\end{equation}

Thus the extended theory (\ref{2.11}) possess solutions that are necessarily confined to spaces with vanishing $^*RR = 2 \partial_\mu K^\mu$. The results of \cite{Jack:2003} indicate that diffeomorphism symmetry breaking effects are barely visible. Parity violation due to Levi-Civita tensor $\epsilon_{\mu\nu\rho}$ causes again the two polarization states to have different intensities. This effect should occur in astronomical situations but milder because of tensor modes of GR.

\section{Conclusions}
\label{conout}

Starting from the beginning of time without singularity we obtain a rather comprehensive picture of the cosmological evolution of the universe from nothing to the present particles. 

Properties of the scenario include:  \\
\noindent
(1) the universe begins in a topological space with the Hartle-Hawking no-boundary condition for the wave function. The high energy density (gravitational or inflaton potential) causes inflationary expansion of spacetime and creation of preon-antipreon pairs. The standard model particles are formed before reheating. Thereafter, the standard model of cosmology takes effect, \\
(2) sparticles have masses on the particle mass scale (this is a problem),  candidate particles are predicted for the dark sector, and the single-family flavor symmetry can be extended to $SU(N_f)$. SUSY breaking remains unsolved,   \\
(3) non-perturbative and perturbatively all order calculable quantum gravity is conjectured. Classicalization is explained by the Wentzel-Kramers-Brillouin (WKB) semiclassical phenomenon and decoherence due to interactions,  \\
(4) mechanism for baryon asymmetry has been constructed,  \\
(5) polarization states of primordial gravitational waves have different intensity. 

The CS action (\ref{eq:nonacs}) is a promising candidate for advancing physics beyond the Standard Model. Remaining items to be studied in this tentative scenario include: the Higgs particle, interactions and masses, SUSY breaking, and in section \ref{qg} the tensor supermultiplet. The main difference between our scenario and the MSSM is compositeness, and the use of the CS action. 

\section*{Acknowledgement}

I thank the two anonymous referees for their careful work and pointing out a number of omissions.

\vskip 2cm
\appendix
\section{Chern-Simons Action}

An instructive introduction to CS theory is found in \cite{Gra:2022}.

The Abelian CS action can be written in terms of $A_{\mu}$ as
\be
S_{CS}[A] = \frac{k}{4\pi} \int_M d^3x\epsilon^{\mu\nu\rho}A_{\mu}\partial_{\nu}A_{\rho} .
\label{eq:nonacs}
\ee

In Euclidean de Sitter gravity, the theory can be expressed using a pair of SU(2) Chern-Simons actions \cite{Ca:2023}

\be\label{eq:dStreeact}
S=k_L\,S_{CS}[A_L]+k_R\,S_{CS}[A_R]~,
\ee
with
\be
S_{CS}[A]=\frac{1}{4\pi}\rm{Tr}\int\left(A\wedge dA+\frac{2}{3}A\wedge A\wedge A\right)~,
\label{eq:Ax2}
\ee
and the trace taken in the fundamental representation. This topological expression is a key element for unification. The other is unbroken supersymmetry.

The gravitational Chern-Simons term $I_{\text{GCS}}$ is
\be
I_\text{GCS}=\frac{1}{2\pi}\rm{Tr}\int\left(\omega\wedge d\omega+\frac{2}{3}\omega\wedge\omega\wedge\omega\right)+\frac{1}{2\pi \ell^2_{dS}} \rm{Tr}\int e\wedge T~,
\ee
with $T$ the torsion two-form and $\ell_{dS}$ is deSitter radius.

\end{document}